# The Distribution of the Energy Gap and Josephson $I_cR_n$ Product in Bi$_2$Sr$_2$CaCu$_2$O$_{8+x}$ by Tunneling Spectroscopy

A. Mourachkine[1]

[1]Université Libre de Bruxelles, CP232, Boulevard du Triomphe, B-1050 Brussels, Belgium

We present direct measurements of the density of states by tunneling spectroscopy on slightly overdoped Bi$_2$Sr$_2$CaCu$_2$O$_{8+x}$ (Bi2212) single crystals at low temperature using break-junction and point-contact techniques. We find that (i) the variation of the gap magnitude, $\Delta$, between 20 and 36 meV is likely to be intrinsic to the Bi2212, and (ii) there is a correlation between the maximum value of the Josephson $I_cR_n$ product and the gap magnitude: $I_cR_n$ decreases with the increase of $\Delta$. The maximum $I_cR_n$ value of 26 mV is observed at $\Delta$ = 20.5 meV. For $\Delta$ = 36.5 meV, the maximum measured value of $I_cR_n$ is 7.3 mV. We conclude that (i) the distribution of the Josephson $I_cR_n$ product as a function of gap magnitude can not be explained by the presence of a single energy gap in Bi2212, and (ii) the coherence energy scale in Bi2212 has the maximum Josephson strength.

**KEY WORDS:** Bi2212; Tunneling; Energy gap; the Josephson product.

## 1. INTRODUCTION

It is known already for some time that, in Bi$_2$Sr$_2$CaCu$_2$O$_{8+x}$ (Bi2212), there is a variation of the magnitude of the energy gap [1-6], which also correlates with the angle into *ab*-plane [1]. This variation is observed in tunneling measurements by STM [2,3,5], break-junctions [4-6] and in planar junctions [1]. The variation of the magnitude of the energy gap has been also observed in tunneling measurements on underdoped Nd$_{1.85}$Ce$_{0.15}$CuO$_{4-\delta}$ (NCCO) [7]. In as-grown Bi2212 single crystals which are slightly overdoped, the gap magnitude, $\Delta$, varies from about 19 meV to 39 meV [1-6]. The $\Delta$ distribution having the $d_{xy}$-like symmetry into *ab*-plane [1] is explained by the presence of two energy scales in cuprates [8,9]. However, there is no consensus whether this variation is intrinsic to Bi2212 or simply because of the variation of the gap magnitude in different samples.

The present work is motivated to shed more light on the question of the "true" superconducting gap in cuprates. There is a consensus that, *at least*, in the underdoped regime of hole-doped cuprates, there are two energy scales: the phase coherence, $\Delta_c$, and pairing, $\Delta_p$, energy scales [8]. By studying the energy gaps with the maximum magnitudes Miyakawa *et al.* showed that, at different hole concentrations in Bi2212, the maximum value of the Josephson $I_c R_n$ product increases with the decrease of hole concentration [10]. Thus, the Josephson product scales with the pairing $\Delta_p$ energy scale indicating that $\Delta_p$ has the predominately superconducting origin [10]. At the same time, the correlation between the Josephson $I_c R_n$ product and magnitude of the energy gap at constant hole concentration has been not studied yet. However, this correlation contains valuable information, and is essential for clarifying the mechanisms behind formation of the superconducting condensate. In the present paper, we present direct measurements of the density of states (DOS) by tunneling spectroscopy on slightly overdoped Bi2212 single crystals at low temperature using break-junction and point-contact techniques. We find that (i) the variation of the gap magnitude approximately between 20 and 36 meV is likely to be intrinsic to Bi2212, and (ii) there is a correlation between the maximum value of the Josephson $I_c R_n$ product and the gap magnitude: The maximum Josephson $I_c R_n$ product decreases with the increase of $\Delta$. This implies that (i) the distribution of the Josephson $I_c R_n$ product as a function of gap magnitude can not be explained by the presence of a single energy gap, and (ii) the coherence energy scale in Bi2212 has the maximum Josephson strength. To our knowledge, this is the first study of the Josephson product as a function of the gap magnitude in Bi2212.

## 2. EXPERIMENTAL

In the present work, we use as-grown Bi2212 single crystals which were grown by using a self-flux method and then mechanically separated from the flux in $Al_2O_3$ or $ZrO_2$ crucibles [6,11]. The dimensions of the samples are typically $2\times1\times0.1$ mm$^3$. The chemical composition of the Bi2212 phase corresponds to the formula $Bi_2Sr_{1.9}CaCu_{1.8}O_{8+x}$ as measured by energy dispersive X-ray fluorescence (EDAX). The crystallographic *a, b, c* values of the single crystals are of 5.41 Å, 5.50 Å and 30.81 Å, respectively. The $T_c$ value was determined by either dc-magnetization or by four-contacts method yielding $T_c$ = 87 - 90 K with the transition width $\Delta T_c \sim 1$ K. The single crystals were checked out to assure that they are indeed overdoped.

Experimental details of our break-junction technique can be found elsewhere [6,11]. Shortly, many break-junctions were prepared by gluing a sample with epoxy on a flexible insulating substrate and then were broken by bending the substrate with a differential screw at low temperature in a helium atmosphere (see the inset in Fig. 1).

The electrical contacts were made by attaching gold wires to a crystal with silver paint. The $I(V)$ and $dI/dV(V)$ tunneling characteristics were determined by the four-terminal method using a standard lock-in modulation technique, which exhibit the characteristic features of typical tunneling spectra in Bi2212, as shown in Fig. 1. During the measurements, by changing the distance between two parts of a broken single crystal it is possible to obtain a few tunneling spectra in one single crystal. In order to confirm the distribution of the gap magnitude obtained in superconductor-insulator-superconductor (SIS) junctions, we also performed tunneling measurements using superconductor-insulator-normal metal (SIN) junctions. Mechanically sharpened Pt-Ir wires were used as a normal-metal tip. The SIN measurements were performed by using the break-junction setup.

The advantage of break-junction technique is that a single crystal is broken *in situ* at low temperature in helium atmosphere. So, the surfaces of the broken crystal are clean, not contaminated. The disadvantage of break-junction technique is that the direction of the tunneling current is not known: the angle in the *ab*-plane can vary from 0 to π, and, even, the tunneling can occur in *c*-direction depending on a break. The magnitude of a superconducting gap can, in fact, be derived directly from the tunneling spectrum. However, in the absence of a generally accepted model for the gap function and the density of states in cuprates, such a quantitative analysis is not straightforward. Thus, in order to compare different spectra, we calculate the gap $2\Delta$ magnitude as a half spacing between the conductance peaks at $V= \pm 2\Delta/e$.

## 3. MEASUREMENTS

The main experimental results of the present study are presented in Figs. 1 - 4. Figure 1 shows $I(V)$ and $dI/dV(V)$ tunneling characteristics measured in a Bi2212 break-junction. The configuration of the break-junction setup used in the measurements is shown schematically in the inset of Fig. 1. In Fig. 1, the conductance curve has a zero-bias conductance peak due to the Josephson current. In our measurements, the Josephson current is present not in all tunneling spectra because of the large value of $R_n$ in some junctions. As we mentioned above, it is possible to obtain a few tunneling spectra in one sample. Figure 2 shows four sets of tunneling spectra, each set is obtained in one Bi2212 sample. The data presented in Figs 2(a)-(c) are obtained in SIS junctions, and Fig. 2(d) shows the data measured in a SIN junction. In Fig. 2, one can see there is a variation of the gap magnitude in each separate sample. Similar variation has been observed in our measurements in 23 single crystals of Bi2212. In Bi2212, we measured a few times tunneling spectra with double-gap structure, similar to that shown in Fig. 2(d). In addition, weak double-gap structures can be also seen in the lower spectrum in the frame (a) and in the upper spectrum in the frame (d) of Fig. 2. The double-gap structures have been also observed in planar junctions [12] and by STM [3]. One can see in Fig. 2 that there is no difference in the distribution of the gap magnitude in SIS and SIN tunneling junctions. However, in SIN junctions, the upper edge of the gap distribution is somewhat larger than that in SIS junctions. This is because in SIN junctions, the surface of a studied crystal is not cleaved *in situ* as in SIS junctions. The gap magnitude measured on uncleaved surfaces is usually somewhat larger than that obtained in SIS junctions [5]. In Fig. 3, we present the statistics of the gap-magnitude measurements in separate single crystals. The variation of the gap magnitude approximately between about 20 and 36 meV is in good agreement with other measurements on as-grown Bi2212 single crystals [1-6]. Thus, we conclude that the variation of the magnitude of the energy gap in Bi2212 single crystals is most likely intrinsic to Bi2212.

One can argue that this variation can be also explained by inhomogeneous oxygen distribution on the surface. This implies that all Bi2212 single crystals studied here and by other groups [1-6,12] are inhomogeneous in the oxygen distribution. It seems that it is not the case, and the variation of the gap magnitude is an intrinsic feature in Bi2212 like in NCCO [7]. We will show further that the maximum Josephson $I_cR_n$ product as a function of the gap magnitude in Bi2212 increases with decrease of the gap magnitude. If we assume that the variation of the gap magnitude in Bi2212 is due to different hole concentration, then, according to Miyakawa *et al.* results [10], the Josephson product has to become larger with increases of the gap magnitude. However, it is not the case. Thus, the distribution of the energy gap in Bi2212 is not due to inhomogeneous oxygen distribution. In fact, there is a physical meaning behind the variation of the gap magnitude in Bi2212 and NCCO since there are two energy scales in cuprates [8] (see Fig. 5).

We turn now to the correlation between the value of the Josephson $I_cR_n$ product and magnitude of the energy gap. Because of the presence of two energy scales in cuprates it is essential to know this correlation (if such exists) for clarifying the mechanism of high-$T_c$ superconductivity. The information about the Josephson current can be only obtained in SIS junctions. In this sense, our work is unique. In addition to this, in the break-junction technique, the studied surfaces of single crystals are clean, so one can relay on the data. Figure 4(a) shows the correlation between the value of the maximum and average Josephson $I_cR_n$ products and the gap magnitude observed in 110 tunneling spectra with the presence of zero-bias conductance peak due to the Josephson current.

The $R_n$ value is estimated from high bias conductance which is relatively constant as shown in Fig. 1. Figure 4(b) shows the statistics of the energy gap for the same 110 tunneling spectra. In Fig. 4(a), one can see that the maximum Josephson product has the maximum of 26 mV at $2\Delta$ = 41 meV. By using the phase diagram of hole-doped cuprates [8] (see Fig. 5), for overdoped Bi2212 with $T_c$ ~ 89 K, we obtain $\Delta_c \approx 21$ meV and $\Delta_p \approx 30$ - 31 meV. This may explain the maximum of the maximum Josephson $I_c R_n$ product at $2\Delta$ = 41 meV, and the hump at $2\Delta$ = 60 meV in the average Josephson product, as shown in Fig. 4(a). The maximum value of 26 mV is in a good agreement with the maximum values of the Josephson product observed in other studies [10,13]. For the energy gap with the maximum magnitude of $2\Delta$ = 73 meV, the maximum $I_c R_n$ value of 7.3 mV is in good agreement with the similar value (7.8 mV) measured by Miyakawa *et al.* [10]. Theoretically, in low-$T_c$ superconductors, $I_c R_n$ is proportional to $\Delta$. From Fig. 4, it is clear that neither the average Josephson $I_c R_n$ product or the maximum Josephson product are in agreement with this proportionality $I_c R_n \sim \Delta$. However, if we assume the presence of more than one gap in Bi2212, then it is possible to explain the obtained results.

From Fig. 4(a), it is clear that the coherence energy scale in Bi2212 has the maximum Josephson strength. Figure 5 shows the phase diagram for a number of cuprates [7,8] and the dependence of the maximum Josephson $I_c R_n$ product on the magnitude of the energy gap. By taking into account the results obtained in the present study and in Ref. 10, we conclude that both the $\Delta_c$ and $\Delta_p$ gaps relate intimately to the superconductivity. This implies that the incoherent Copper pairs exist above $T_c$ and condense into the coherent state at $T_c$ [8,11].

Finally, by taking into account the angular dependence of the magnitude of the energy gap in Bi2212, which reminds the $d_{xy}$- like symmetry into *ab*-plane [1], and Fig. 4 (a), the $I_c R_n(\theta)$ dependence, where $\theta$ is the angle into *ab*-plane, would have the $d_{x^2-y^2}$-like symmetry. Indeed, there is a consensus that the predominant order parameter in hole-doped cuprates has the $d_{x^2-y^2}$ symmetry [14].

## 4. CONCLUSIONS

In summary, we presented direct measurements of the density of states by tunneling spectroscopy on slightly overdoped Bi2212 single crystals at low temperature using a break-junction technique. We found that (i) the variation of the gap magnitude between 20 and 36 meV is likely to be intrinsic to the Bi2212, and (ii) there is a correlation between the maximum value of the Josephson $I_c R_n$ product and the gap magnitude: The maximum Josephson $I_c R_n$ product decreases with the increase of $\Delta$.

The maximum $I_cR_n$ value of 26 mV is observed at $\Delta$ = 20.5 meV. For the gap with the maximum magnitude ($\Delta$ = 36.5 meV), the maximum measured value of $I_cR_n$ is of 7.3 mV. New and central results obtained in this study are that (i) the distribution of the Josephson $I_cR_n$ product as a function of gap magnitude, neither the average nor maximum Josephson $I_cR_n$ product, can not be explained by the presence of a single energy gap in Bi2212, and (ii) the coherence energy scale in Bi2212 has the maximum Josephson strength.

**REFERENCES**

1. J. Kane and K.-W. Ng, *Phys. Rev. B* **53**, 2819 (1996).
2. A. Chang *et el.*, *Phys. Rev. B* **46**, 5692 (1992).
3. Ch. Renner and O. Fisher, *Proc. SPIE* **2158**, 135 (1994).
4. S. I. Vedeneev, A. G. M. Jansen, and P. Wyder, *Physica B* **218**, 213 (1996).
5. H. Hancotte *et al.*, *Phys. Rev. B* **55**, R3410 (1997).
6. A. Mourachkine, *J. Superconductivity* **13**, 101 (2000).
7. A. Mourachkine, *Europhys. Lett.* **50**, 663 (2000).
8. G. Deutscher, *Nature* **397**, 410 (1999).
9. A. Mourachkine, *Physica C* **323**, 137 (1999).
10. N. Miyakawa et al., *Phys. Rev. Lett.* **83**, 1018 (1999).
11. A. Mourachkine, *Europhys. Lett.* **49**, 86 (2000).
12. H. J. Tao, F. Lu, and E. L. Wolf, *J. Phys. Chem. Solids* **52**, 1481 (1991).
13. H. J. Tao, F. Lu, G. Zhang, and E. L. Wolf, *Physica C* **224**, 117 (1994).
14. D. J. Van Harlingen, *Rev. Mod. Phys.* **67**, 515 (1995).

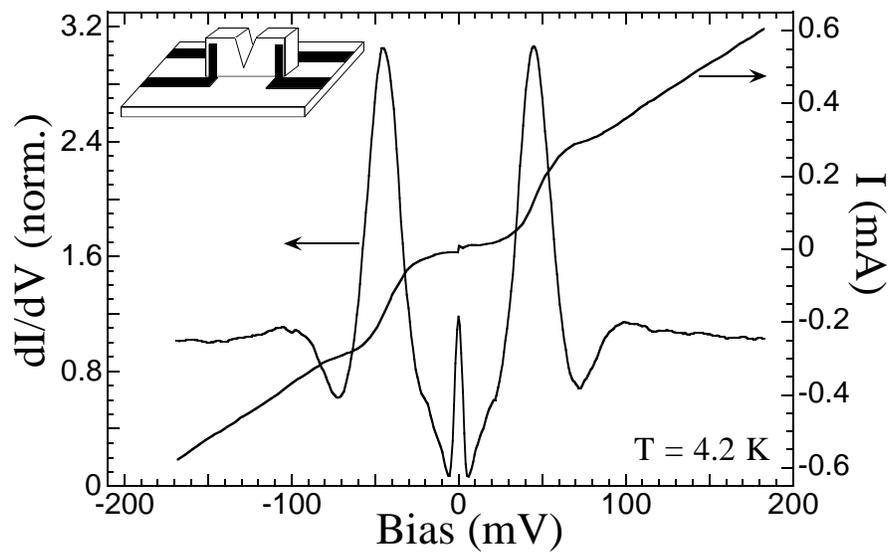

**Fig. 1.** *I(V)* and *dI/dV(V)* tunneling characteristics measured in a Bi2212 break-junction. The inset shows schematically the sample after mounting on a flexible substrate. The gold wires are shown schematically by the black stripes.

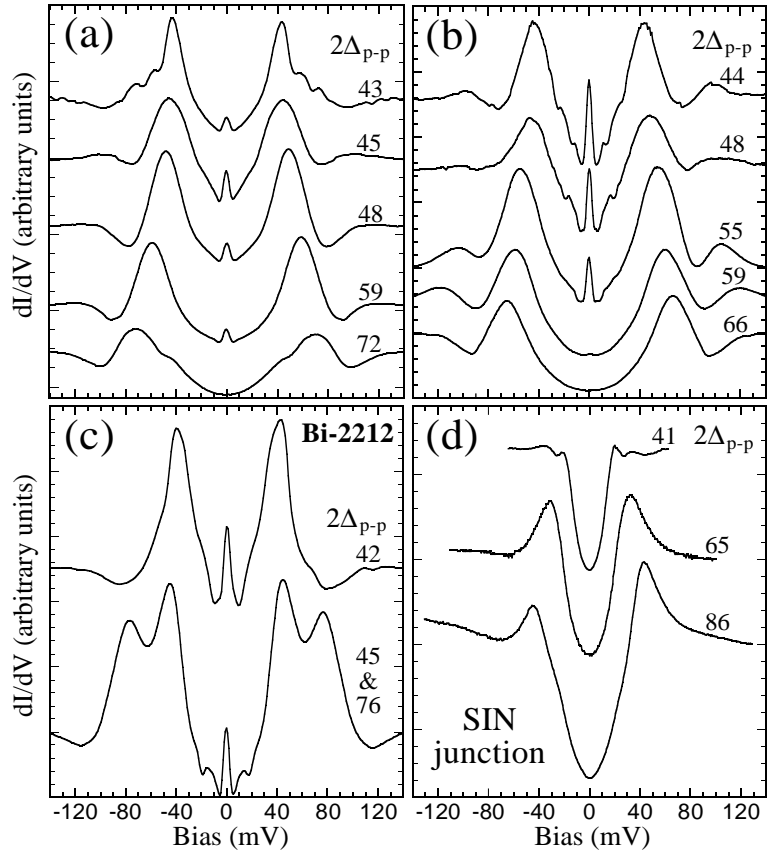

**Fig. 2.** *dI/dV(V)* tunneling characteristics measured in Bi2212 in SIS junctions (a)-(c) and in a SIN junction (d). The spectra in each frame are obtained in one single crystal. In each frame, the spectra are offset vertically for clarity. The peak-to-peak values, $2\Delta_{p-p}$, are presented in meV. In frame (c), the lower spectrum has the double-peak structure. In all frames, one can see that the magnitude of the energy gap is not constant, but varies.

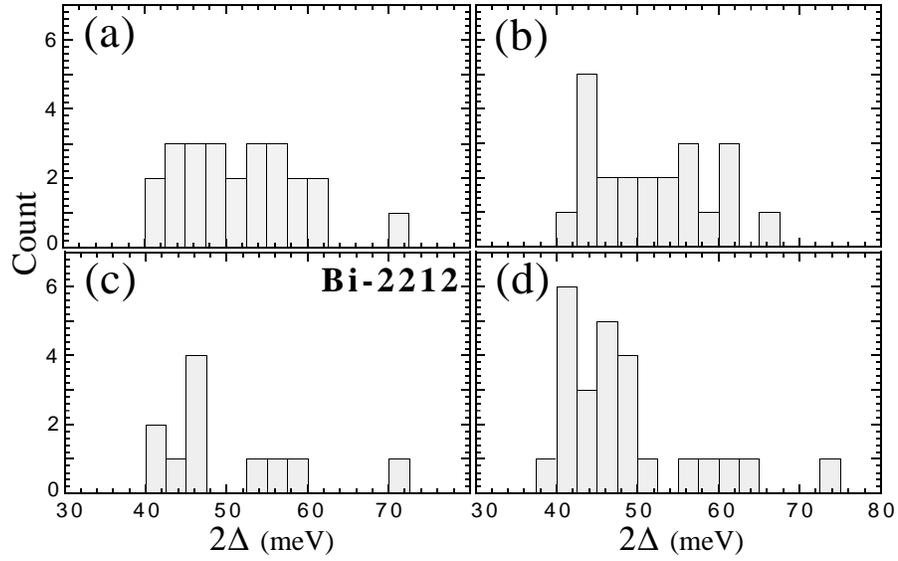

**Fig. 3.** (a)-(d) Statistics of the magnitude of the energy gap, $2\Delta_{p\text{-}p}$, in different Bi2212 break junctions. Each histogram is obtained in one single crystal. In all histograms, one can see that the magnitude of the energy gap is not constant, but varies.

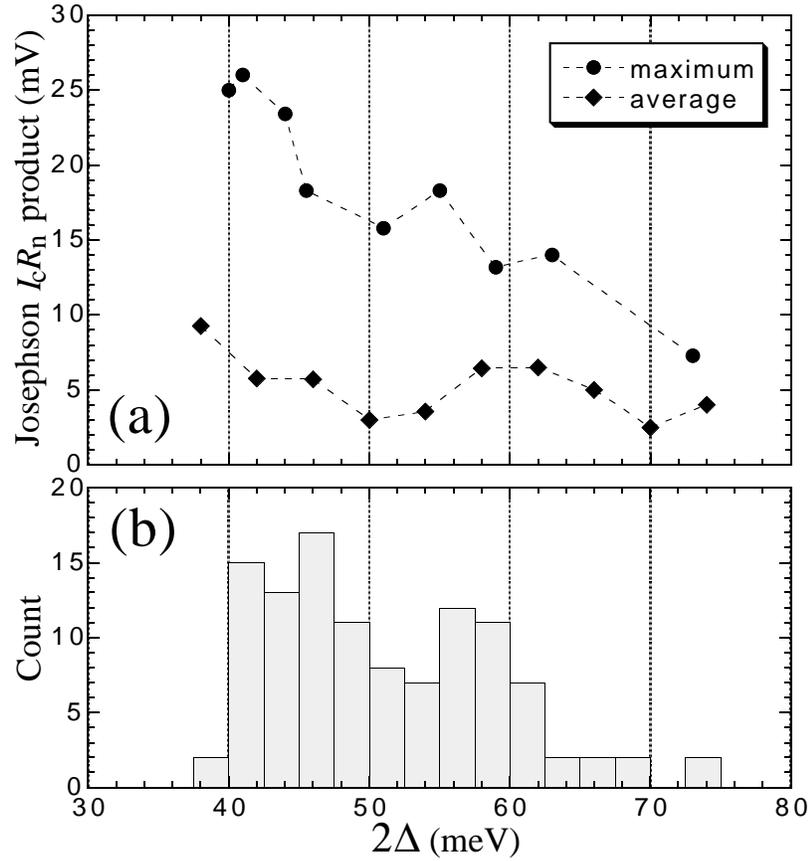

**Fig. 4.** (a) Maximum and average Josephson $I_cR_n$ product as a function of the gap magnitude in 110 Bi2212 break junctions. The maximum value of the maximum Josephson product (26 mV) is measured at $2\Delta = 41$ meV. (b) Statistics of the magnitude of the energy gap for the same 110 junctions.

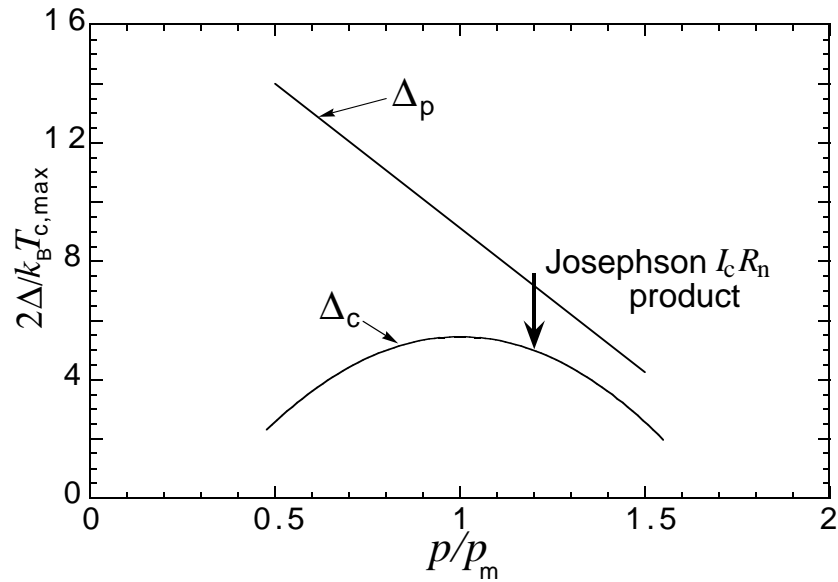

**Fig. 5.** Phase diagram of hole-doped cuprates at low temperature: $\Delta_p$ is the pairing energy scale, and $\Delta_c$ is the phase coherence scale [8]. Here $p_m$ is a hole concentration with the maximum $T_c$. The thick arrow shows the direction in which the maximum Josephson $I_c R_n$ product increases (present work). The point $p/p_m = 1.2$ corresponds approximately to overdoped Bi2212 with $T_c \sim 89$ K [6].